\documentclass{article}
\usepackage{graphicx}
\begin{document}
\title{Semiclassical description of electronic \\
supershells in simple metal clusters}
\author{G.V.Shpatakovskaya\\
{\it Institute for Mathematical Modelling, RAS}\\
{\it 125047, Russia, Moscow}}
\maketitle
\begin{abstract}
A semiclassical approach for calculating shell effects, that has
been used  in atomic and plasma physics, is applied to describe electronic
supershells in metal clusters. Using the spherical jellium model
we give the analytical expression for the oscillating part of the binding
energy of electrons as an explicit sum of contributions from supershells with 
quantum numbers $2n_r+l$, $3n_r+l$, $4n_r+l$, ... This expression is written
in terms of the classical characteristics of the motion of an electron with 
the Fermi energy	in a self-consistent potential. The conditions under which 
a new supershell appears  and the relative contribution of this shell are studied
as a function of the cluster size and form of the potential. Specific calculations
are performed for a square well.
\end{abstract}

  {\bf 1.} The electronic structure of atomic clusters has been intensively studied 
experimentally and theoretically in the last two decades (see, for example, reviews 
in Refs. \cite{Heer-93} and \cite{Brack-93}). One general feature of the experimental 
mass-abundance spectra of $N$-atom clasters is the existence of magic numbers:
Clusters with these numbers of atoms are produced more abundantly. As $N$ enlarges, 
the amplitude of these variations lessens, then in still larger clusters increases 
once again, and so on, i.e., oscillations with beats occur.

  The theoretical reproduction of the beating patterns one can see,
for example, in the self-consistent calculations of sodium clusters in Ref.
\cite{Genzken-92}, where the spherical jellium model in local
density approximation was used. There are two kinds of periodicity 
in the density of states and in the oscillating part $\delta E_{sh}$
of the binding energy as a function of a cluster radius	$R \sim	N_{e}^{1/3}$
($N_{e}$ is a number of valence electrons in cluster):
the primary shell periodic structure and beat mode with a higher-order
period. The similar results were performed in Ref. \cite{Nishioka-90} using the
Woods-Saxon potential and in calculations for various metals in Ref.
 \cite{Clemenger-91} using nearly self-consistent potentials.

	For small clusters  the interpretation of the numerical results is
obvious: the extremum energy cusps occur at	$l_{max}$-shell closing and
represent the magic numbers ($l_{max}$ is a maximum orbital number $l$
in the cluster). However this law breaks down for $N_e > 100$.

  The theory, explaining quantitatively the beating pattern of the level
density for spherical cavity by a superposition of the contributions from 
closed classical trajectories of electrons, has been developed in the fundumental work
of Balian and Bloch \cite{Balian-71}. The detailed
numerical calculations in Ref. \cite{Clemenger-91} using more complex spherical
cluster potentials reproduce the similar oscillations, one-electron levels
$\varepsilon(n_r,l)$ of the higher angular momentum states bunching in
supershells: $\varepsilon(n_r,l)\simeq\varepsilon(n_r+1,l-K)$ with
pseudoquantum numbers $K n_r+l$. Here $ n_r$ is a radial quantum number,
$K = 2,3,4,...$. The integer	number $K$ under classical treatment (see, for
example in Ref.\cite{Bjornholm-92}) is equal to a ratio of the frequencies in the 
radial and angular motion of the corresponding closed	 orbit.
 At $K = 2$ pseudoquantum number is the same as a principal quantum number
and characterizes the pendulating orbit going through the origin, $K = 3$
and $K = 4$ corresponds to the triangular and square orbits, respectively.

One can suppose from here that a periodic-orbit-expansion (or supershell-
expansion), obtained for a spherical cavity, is a particular case of the
more general expansion for spherical cluster potentials. This conclusion
is supported in the recent paper \cite{Koch-98} where such a generalization
was carried out using Woods-Saxon potential by expanding on a parameter $a/R$
($a$ is a surface width) around the known results for a spherical potential
well. It is of interest to investigate analytically the origin of supershells and
the mechanisms leading to their appearance for a potential of arbitrary form.

	In the present paper it is shown that this problem can be solved by a 
semiclassical method for distinguishing shell effects, previously applied 
successfully in atomic \cite{Kirzhnits-75},\cite{Englert-85}, and plasma 
\cite{Shpatakovskaya-85},\cite{Kuzmenkov-92} physics on the basis of the Thomas-Fermi
(TF) model. Although the TF model and its conventional variants ETF with quantum and 
exchange corrections (for application to clusters, see, for example,Refs. 
\cite{Kresin-88} and \cite{Membrado-90}) give only the average dependences of all 
quantities on the number of particles, a refinement of this model makes it possible 
to account for the shell structure of the electronic spectrum. This refinement is 
based on the use of the Bohr-Zommerfeld quantization conditions and on the 
possibillity of performing the sum over quantum numbers analytically, provided that 
the semiclassity parameter, which for clusters is proportional to $ N_{e}^{-1/3}$, 
is small.

  {\bf 2.} We furthermore use the expression derived in our future extended publication
for the correction to the binding energy of electrons (the atomic units are used):
\begin{equation}
\delta E=
\int\limits_{-\infty}^{\mu}d\mu '\int d{\bf{r}}\delta n(\bf{r},\mu ')
\end{equation}
Here $\delta n(\bf{r})$ is the correction to the electronic density, which because
of other effects goes beyond the initial model approach (for example, TF or ETF model),
$\mu$ is the chemical potential in the initial model, the correction $\delta n(\bf{r})$
is assumed to be small and is calculated using the initial self-consistent potential.

We are interested in the contribution of the shell correction to the electronic
density of states $\delta n_{sh}(\bf{r},\mu)$ or to the number of states:
\begin{equation}
\delta N_{sh}(\mu)=\int d{\bf{r}}\delta n_{sh}({\bf{r}},\mu)
					 =N(\mu)-N_{TF}(\mu),
\end{equation}
where for a cluster with the closing "$l$"-shells 
\begin{equation}
 N(\mu ) = 2\sum_{n_r, l}(2l+1) \theta (\mu - \varepsilon_{n_r ,l}),
\end{equation}
and in the semiclassical approach the energy levels $\varepsilon_{n_r ,l} $ are 
determined from the quantization condition 
\begin{equation}
S_{\varepsilon l}=\int dr p_{\varepsilon l}(r)=\pi\left(n_r +\frac{1}{2}\right)
\end{equation}
Here $ S_{\varepsilon l}\equiv S_{\varepsilon\lambda}$ and $p_{\varepsilon l}(r)=
\sqrt{2(\varepsilon-U(r))-(l+1/2)^2/r^2}\equiv
\sqrt{p_\varepsilon^2(r)-\lambda ^2/r^2}\equiv p_{\varepsilon\lambda}$ are, 
respectively,
the classical radial action and momentum of an electron with energy $\varepsilon$ 
and orbital angular momentum $l$. Simple calculations using the Poisson formula to
replace the sums over quantum numbers $n_r$ and $l$ by integrals make it possible
to rewrite expression (3) as
\begin{equation}
 N(\mu )=\frac{2}{\pi}\sum_{k,s=-\infty}^{\infty}\frac{(-1)^{k+s}}{k}
\int\limits_{0}^{\lambda_{\mu}}d\lambda\lambda
\sin (2\pi k\nu_{\mu\lambda})\cos (2\pi s\lambda)
\end{equation}
Here $\nu_{\varepsilon\lambda}=S_{\varepsilon\lambda}/\pi$, and 
$\lambda_{\varepsilon}$ determines the border of the phase area of the
classically allowed motion of an electron with an energy $\varepsilon$:
$\nu_{\varepsilon\lambda_\varepsilon} =0$. In Eq.(5) the term with $k=s=0$ gives 
the TF result $N_{TF}(\mu)$ and according to Eq.(2) the sum (5) without 
this term is equal to the desired quantity $\delta N_{sh}(\mu)$.

Let's note now, that the supposed approach is agreed with a general, known from
the nuclear physics, concept \cite{Strutinsky-68},\cite{Yannouleas-93} of separating 
the total energy of finite sistem as a function of the system size into a smooth part 
and a fluctuating correction. But here for calculating the last we base the simple 
expression (1), using initial smooth self-consistent potential or its Woods-Saxon 
and another fitting.

 {\bf 3.} The integral limits and points $\bar{\lambda}$ of stationary phase make 
the main contribution in the integral over $\lambda$ in Eq.(5). These points are 
determined from the relation:
\begin{equation}
\frac{\partial\nu_{\mu\lambda}}{\partial\lambda}\Bigl|_{\bar{\lambda}}=-\frac{s}{k},
\qquad  0\leq\lambda\leq\lambda_{\mu}
\end{equation}
A function $\nu_{\mu\lambda}(\lambda)$ decreases monotonically  and for all potentials 
$U(r)$ that are finite at the origin the slope of the corresponding curve at 
$\lambda = 0$ is the same \cite{Kirzhnits-72}, 
$$
\frac{\partial\nu_{\mu\lambda}}{\partial\lambda}\Bigl|_0=-\frac{1}{2},
\eqno (6a)$$
The value of the derivative at $\lambda=\lambda_{\mu}$
$$
\frac{\partial\nu_{\mu\lambda}}
{\partial\lambda}\Bigl|_{\lambda_{\mu}}\equiv-\nu_{\mu}'
\eqno (6b)$$
depends strongly on the form of the potential. For an oscillator  
$\nu_{\mu}'=1/2$, for a square well $\nu_{\mu}'=0$, and for the Woods-Saxon potential 
the value of $\nu_{\mu}'$ varies with increasing $N$, vanishing in the limit of a very
large number of atoms.

   On this basis it follows that the relation (6) distinguishes in the sum over $k$ 
the leading terms   
\begin{eqnarray}
k=(2+j)s,\quad
\frac{\partial\nu_{\mu\lambda}}{\partial\lambda}\Bigl|_{\bar{\lambda}_j}
=-\frac{1}{2+j},\nonumber\\ j=0,1,...j_{max},\quad
 j_{max}=\left[\frac{1}{\nu_{\mu}'}-2\right]
\end{eqnarray}
The terms $k=2s (j=0)$ must be studied separately,	 since in this case the point of
stationary phase $\bar{\lambda}=0$ is also the lower limit of integration. As 
a result we obtain
\begin{eqnarray}
\delta N_{sh}=\sum_{s=1}^{\infty}
\frac{(-1)^s}{(\pi s)^2}
\left\{
\frac{\cos(2\pi s2\nu_{\mu 0})}{\delta_{\mu}^{(0)}}-
\frac{\lambda_{\mu}\cos (2\pi s\lambda_{\mu})}{0.5-\nu_{\mu}'}-
\right.\nonumber\\
\left.-\sum_{j=1}^{j_{max}}\frac{4\sqrt{s}\cdot j\cdot (-1)^{j\cdot s}}
{(\delta_{\mu}^{(j)})^{3/2}(2+j)^{5/2}}\cos
\left[
2\pi s
\left(
(2+j)\nu_{\mu\bar{\lambda}_j}+\bar{\lambda}_j
\right)
-\frac{\pi}{4}
\right]
\right\}
\end{eqnarray}
Here
$$\delta_{\mu}^{(j)}\equiv\frac{\partial^2\nu_{\mu\lambda}}{\partial\lambda^2}
\Bigl|_{\bar{\lambda}_j}.$$ 
Substituting expression (8) into Eq.(1) and integrating by parts to separate the terms
which are of leading order in the semiclassicity parameter give a semiclassical formula for the 
shell correction to the binding energy of electrons in a cluster 
\begin{eqnarray}
\delta E_{sh}=\frac{1}{2}\sum_{s=1}^{\infty}
\frac{(-1)^s}{(\pi s)^3}
\left\{
\frac{\sin(2\pi s2\nu_{\mu 0})}
{\delta_{\mu}^{(0)}(\partial 2\nu_{\mu 0}/\partial\mu)}
-\frac{\lambda_{\mu}
\sin (2\pi s\lambda_{\mu})}{\left( 0.5-\nu_{\mu}'\right)
(\partial\lambda_{\mu }/\partial\mu)}\right.-
\nonumber\\
\left.-\sum_{j=1}^{j_{max}}\frac{4\sqrt{s}\cdot j\cdot (-1)^{j\cdot s}}
{(\delta_{\mu}^{(j)})^{3/2}(2+j)^{5/2}}\frac
{\sin\left[2\pi s\left((2+j)\nu_{\mu\bar{\lambda}_j}+\bar{\lambda}_j\right)
-\frac{\pi}{4}\right]}{ \partial\left( (2+j)\nu_{\mu\bar{\lambda}_j}+
\bar{\lambda}_j \right)/\partial\mu }
\right\},
\end{eqnarray}
as an sum of contributions from supershells with quantum numbers
$n_j = K n_r+l,\quad K=2+j$, the quantization on the Fermi level being substantial.
In Eqs.(8) and (9) the symbol $j$ numerates a kind of the electronic orbit: $j=0$
corresponds to the linear pendulating orbit, the terms $j \geq 1 \quad (K\geq 3) $ 
are connected with the planar regular polygons, $K$ being the number of their
vertices. The integer value $s$ is equal to a number of periods that the electronic
trajectory $(j,s)$ 
\twocolumn
[includes, so the sum over $s$ is the trajectory
length-expansion for a $j$-orbit (compare with the orbits $(\lambda,\nu)$ in Ref.
\cite{Koch-98}).

\hspace{0.5cm}{\bf 4.} The proposed simple method makes it possible for any
spherical potential to determine the period and amplitude of oscillations associated with each 
supershell and to estimate their relative role in beating occurence. The results 
of such an analysis for a square well potential: 
$$
U(r)=\left\{ -2\varepsilon_F, r\leq R\atop 0,r > R\right. ,\quad
R=r_s N_{e}^{1/3},
\quad \varepsilon_F= \frac{1}{2r^2_s}\left( \frac{9\pi}{4}\right)^{2/3},
\quad \mu = - \varepsilon_F 
$$
are displayed in Fig.1. Let's note that even though we start with the expression] 
\noindent(3) for a cluster with the closing "$l$"-shells our results	 for $N_e > 100$
check well with the results of the complete calculations in Ref. \cite{Clemenger-91}.
\begin{figure}[p]

\includegraphics[0.0in,0in][6cm, 11cm]{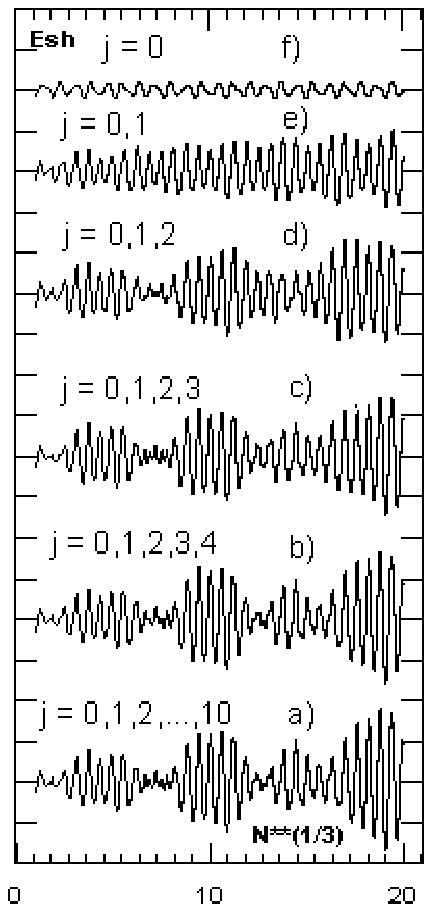}

\caption{ a) -- Shell energy correction $\delta E_{sh}$ as a function of $N_{e}^{1/3}$.
The \mbox{unit of energy is the Fermi energy $\varepsilon_F$.}
\mbox{b)-f) -- Comparasion of a truncated} supershell-expansions.}
\end{figure}
Fig.1 shows, that the first period \mbox{($N_{e}^{1/3}<7$)} of the beats is determined
by terms with $j=0,1,2$, the term with $j=0$ contributing little. The account for term 
with $j=3$ ($K=5$) is needed to describe well the second period 
\mbox{($N_{e}^{1/3}\leq 13$)}. Add- ing a term with $j=4$ ($K=6$) is sufficient to 
describe the behavior of $\delta E_{sh}$ in the entire range under study. This means 
that the actual value $j_{max}$ is less that determined from Eq.(7) and corresponds to
the filled states, for which $$\nu_{\mu\bar{\lambda}_j}\geq\frac{1}{2}.$${\sloppy

} 
\hspace{0.25cm}{\bf 5.} For spherical potential we give the analytical 
expression for the oscillating part of the binding energy of electrons as an explicit 
sum of contributions from the leading	periodic orbits (supershells).
We show that a number of the leading 
orbits in the expansion depends strongly on the form of the potential 
and is determined by the value of derivative (6b)(see Eq.(7)).
 The more soft potentials will be dealt in detail in our
extended publication where among other things small clusters and finite temperatures 
will be analyzed.{\sloppy

} 
\onecolumn

\end{document}